\documentclass[12pt]{iopart}
\usepackage[utf8x]{inputenc}
\usepackage[T1]{fontenc}
\usepackage{iopams}  
\usepackage{subfigure}
\usepackage{graphicx}
\usepackage{array}
\usepackage{xcolor}
\usepackage{booktabs}
\usepackage{amssymb}
\usepackage{hyperref}
\expandafter\let\csname equation*\endcsname\relax
\expandafter\let\csname endequation*\endcsname\relax 
\usepackage{mathtools} 
\graphicspath{{./Figures/}}
\usepackage[symbol]{footmisc}

\begin{document}

\title[]{Gravitational redshift test with the future ACES mission}

\author{E~Savalle$^{1,\footnotemark[2]}$,  C Guerlin$^{1,2,\footnotemark[2]}$, P~Delva$^1$, F~Meynadier$^{1,3}$, C~le~Poncin-Lafitte$^1$,
P~Wolf$^1$ \footnotetext {E. Savalle and C. Guerlin should be considered as co-first authors.}} \address{$^1$SYRTE, Observatoire de Paris, Universit\'e PSL, CNRS, Sorbonne Universit\'e, LNE, 61 avenue de
l'Observatoire, 75014 Paris, France} \address{$^2$Laboratoire Kastler
Brossel, ENS-Université PSL, CNRS,
Sorbonne Universit\'e, Coll\`ege de
France, 24 rue Lhomond, 75005 Paris, France}  \address{$^3$Bureau International des Poids et Mesures, Pavillon de Breteuil, 92312 S\`{e}vres Cedex, France}

\ead{christine.guerlin@upmc.fr}
\vspace{10pt}
\begin{indented}
\item[]25th June, 2019
\end{indented}

\begin{abstract}
We investigate the performance of the upcoming ACES (Atomic Clock Ensemble in Space) space mission in terms of its primary scientific objective, the test of the gravitational redshift. Whilst the ultimate performance of that test is determined by the systematic uncertainty of the on-board clock at 2-3 ppm, we determine whether, and under which conditions, that limit can be reached in the presence of coloured realistic noise, data gaps and orbit determination uncertainties. To do so we have developed several methods and software tools to simulate and analyse ACES data. Using those we find that the target uncertainty of 2-3 ppm can be reached after only a few measurement sessions of 10-20 days each, with a relatively modest requirement on orbit determination of $\sim$300~m.  
\end{abstract}
\maketitle
%
%
%
%
%

\section{Introduction}

General relativity~(GR) together with all other metric theories of
gravitation provides a geometrical description of the gravitational
interaction. Fundamentally, such a description is based on the
Einstein equivalence principle~(EEP), itself the result of universal
coupling of all standard matter to gravity. Although very successful
so far, there are reasons to think that sufficiently sensitive
measurements could uncover a failure of the EEP. For example, the
unification of gravitation with the other fundamental interactions,
and quantum theories of gravitation, generally lead to small
deviations from the EEP~(see \textit{e.g.}~\cite{Will2018}). Also dark matter
and energy are so far only observed through their gravitational
effects, but might be hints towards a modification of GR.

From a phenomenological point of view, three aspects of the EEP can be
tested: (i) the universality of free fall~(UFF); (ii) local Lorentz
invariance~(LLI); and (iii) local position invariance~(LPI). Although
the three are related, the quantitative details of that relation are
model dependent \cite{Will2018,Wolf2016}, so each of the three
sub-principles needs to be tested independently to best possible
uncertainty. UFF has been recently constrained by the Microscope space
mission~\cite{Touboul2017}, while LLI was recently constrained, for
example, by using a ground fibre network of optical
clocks~\cite{Delva2017e} (see \textit{e.g.}~\cite{Will2018,Mattingly2005} for
reviews). In this paper we focus on testing LPI.

LPI stipulates that the outcome of any local non-gravitational
experiment is independent of the space-time position of the
freely-falling reference frame in which it is performed. This
principle is mainly tested by two types of experiments: (i) search for
variations in the constants of Nature (see \textit{e.g.}~\cite{Uzan2011} for a
review) and (ii) gravitational redshift tests. The gravitational
redshift was observed for the first time in the Pound-Rebka-Snider
experiment~\cite{Pound1960,Pound1959,Pound1959a,Pound1965}.

One of the most accurate tests of the gravitational redshift has been
realized with the Vessot-Levine rocket experiment in 1976, also named
the Gravity Probe A (GP-A)
experiment~\cite{Vessot1980,Vessot1979,Vessot1989}. The frequency
difference between a space-borne hydrogen maser clock and ground
hydrogen masers was measured thanks to a continuous two-way microwave
link. The total duration of the experiment was limited to 2 hours and
reached an uncertainty of $1.4\times 10^{-4}$~\cite{Vessot1989}. Very
recently this has been surpassed by the analysis of clocks on board
two eccentric Galileo satellites, reaching an uncertainty of
$2.5\times 10^{-5}$ \cite{Delva2015,Delva2018,Herrmann2018}.

In this work we study the expected performance of the gravitational
redshift test using the ACES (Atomic Clock Ensemble in Space) mission,
scheduled for launch in 2020. The heart of that mission is an accurate
cold atom Cesium clock (PHARAO). It will be installed on an outside pallet
of the Columbus module of the international space station (ISS) at 400
km altitude. Comparing that clock to ground clocks using a two-way
microwave link (MWL) will allow performing a redshift test at an
expected uncertainty of $2-3\times 10^{-6}$, limited by the systematic
uncertainty of PHARAO. Here we investigate whether, and under which
conditions, that goal can be reached when taking into account the main
statistical noise contributions and the uncertainty from orbit
determination errors.

The MWL and its data analysis has been described in
\cite{Meynadier2018}, which also provides an estimate of the effect of
orbit determination errors on the time transfer model, but not on the 
determination  of the frequency shift of the clock. The latter has been 
studied to some extent in \cite{Duchayne2009}. Here we study, in a full 
end to end scenario, the contribution of all noise sources (MWL, clocks), 
as well as the orbit determination errors, on the final scientific goal,
\textit{i.e.} the test of the gravitational redshift. To do so we use two 
software tools developed specifically for the ACES mission that simulate 
and analyse the data. The latter will be used for ACES data analysis when
the mission will fly.

Our paper is organized as follows: Section \ref{sec:ACES_mission}
provides a brief description of the ACES space mission, its payload,
and its specifications, together with a description of the
gravitational redshift in that context. We conclude that section by an
overview of our simulation and analysis software. Data simulation and
analysis are then presented in details respectively in Sections \ref{simulation} and
\ref{modeling}. We describe different data observables and
corresponding models, with an emphasis on parameters estimation in
the presence of realistic noise and data gaps. We then provide our
results concerning the gravitational redshift test accuracy (Section
\ref{redshift}) and the required uncertainty on orbit determination
(Section \ref{orbito}). Finally we conclude with a summary of our main
results and a view towards the future in Section \ref{sec:Conclusion}.

\section{The ACES mission and the gravitational redshift} \label{sec:ACES_mission}

\subsection{Overview}
During 18 months up to 3 years, the ACES module will be attached to the ISS
and up to 8 ground terminals (GT) will operate in ground laboratories
\cite{Meynadier2018,Cacciapuoti2009, TN, Laurent2015}. A
two-way microwave link (MWL) will allow time comparison between ground clocks and
the onboard timescale provided by the cold atom clock PHARAO and a
hydrogen maser (SHM). Ground to space comparisons will be made when the ISS is in view of a given
GT. The phase accumulated by the clocks between passes of the ISS is kept track of and
allows the monitoring of the desynchronisation coherently over the typical duration of
continuous operation of PHARAO and grounds clocks, which will be
limited by the ISS environment (manoeuvring and other disturbances) to periods of typically 10-20 days. The primary scientific objective of the mission is to
measure Einstein's gravitational redshift at 2-3 ppm, improving the present best test \cite{Delva2018,Herrmann2018} by about a factor 10. 

\subsection{Einstein's gravitational redshift}

According to GR, the proper time $\tau$ of a
clock near the Earth evolves as:
\begin{equation} \frac{d\tau}{dt}= 1-\left(\frac{U(t,{\bf
x})}{c^2}+\frac{{\bf v}(t)^2}{2c^2}\right)+ \Or(c^{-4})
\label{eq-dtaudt}
\end{equation}
\noindent where $c$ is the speed of light in vacuum, $(t,{\bf x})$
are the space-time coordinates in a geocentric non-rotating coordinate system
(the Geocentric Coordinate Reference System, GCRS \cite{Soffel2003}), ${\bf v}(t)$ is
the clock coordinate velocity in GCRS at coordinate time $t$, and
$U(t,{\bf x})$ is the Newtonian potential at time $t$ and position
${\bf x}$ (taken with the convention $U>0$). This expression is valid
in the post-newtonian approximation, for low potential ($U/c^2\ll1$)
and velocity ($v^2/c^2\ll1$). Higher order terms in (\ref{eq-dtaudt})
have been investigated \textit{e.g.} in \cite{Wolf1995, Petit2005} and are
negligible at the sensitivity of ACES. Other higher order terms of
order $\Or(c^{-3})$ can play a role in frequency transfer
{\cite{Blanchet2001}} at $\sim 10^{-16}$ uncertainty in fractional
frequency, but are negligible for the time transfer used in ACES.

In Eq. (\ref{eq-dtaudt}) the first term is the gravitational redshift,
the second term is the second order Doppler effect from Special
Relativity. The expression (\ref{eq-dtaudt}) is also equal to the
frequency ratio between two clocks:
\begin{equation}
\frac{d\tau}{dt}=\frac{\nu}{\nu_0} 
\end{equation}
where $\nu_0$ (resp. $\nu$) is the frequency of a clock at rest and at zero
gravitational potential (resp. at non-zero gravitational potential and
velocity). We define the fractional frequency shift between these two clocks as:
\begin{equation}
y(t)=\frac{d\tau}{dt}-1=\frac{\Delta\nu}{\nu_0} 
\label{eq-ydef}
\end{equation}
\noindent where $\Delta\nu=\nu-\nu_0$.
 Clocks at non-zero gravitational potential
tick slower and are thus ``red-shifted'' by the gravitational
potential. 

The gravitational redshift can be measured by
comparing two clocks ($g$ and $s$) at different gravitational potentials, \textit{i.e.} by determining $y_s(t)-y_g(t)$. For a
clock on ground in Paris ($g$), the gravitational redshift is
$-U_g/c^2=-6.96\times10^{-10}$. For a space clock ($s$) on board the ISS (height
above Paris
$\approx$~350~km), the gravitational redshift is
$-U_s/c^2=-6.60\times10^{-10}$. Thus PHARAO, considering only this term, ticks faster with a
differential gravitational redshift of:
\begin{equation}
\label{eq-redshift}
-\frac{\Delta U}{c^2}=-\frac{U_s-U_g}{c^2}\approx3.6\times10^{-11}.
\end{equation}
\noindent The overall rate of PHARAO is actually slower than a static ground clock due to the
bigger contribution of the Doppler effect term, for which the difference
has opposite sign and is:
\begin{equation}
\label{eq-doppler}
-\frac{\Delta{\bf v}^2}{2c^2}=-\frac{{\bf v}_s^2-{\bf v}_g^2}{2c^2}\approx -3.3\times10^{-10}.
\end{equation}

Comparing the two clocks can be realized by exchanging electromagnetic signals, either by frequency
comparison, or time comparison if one is able to monitor their
desynchronization
$\Delta\tau(t)=\tau_s(t)-\tau_g(t)$. Eq. (\ref{eq-redshift}) gives the
contribution of the gravitational redshift term to the
differential relative frequency shift. Their overall desynchronization can be
obtained by integrating Eq. (\ref{eq-dtaudt}) :
\begin{equation}
\label{eq-deltatau}
\Delta \tau(t)=\Delta\tau_0-\int_{t_0}^t\frac{\Delta
  U(t')}{c^2}dt'-\int_{t_0}^t\frac{\Delta{\bf v}^2 (t')}{2c^2}dt'
\end{equation}

\noindent where $\Delta\tau_0=\Delta\tau(t_0)$ is the initial phase offset between the two clocks. For a constant potential difference as given in
Eq. (\ref{eq-redshift}), the gravitational redshift thus leads to a linear drift of the
desynchronization by about 3
$\mu$s per day.

A deviation from the gravitational redshift of General Relativity is commonly
sought as a constant and isotropic deviation, where Einstein's
gravitational redshift is rescaled by a factor $1+\alpha$, with
$\alpha=0$ for GR \cite{Will2018, Wolf2016}. 
Assuming the measurement is limited only by the systematic effects on PHARAO frequency at $\delta y \leq 1\times10^{-16}$, the sensitivity of the gravitational redshift test would reach an uncertainty on $\alpha$ of about 2-3 ppm, improving by one order of magnitude the stringent test achieved using the accidental eccentricity of two Galileo satellites \cite{Delva2015,Delva2018, Herrmann2018}.

The aim of this work is to investigate whether, and under which
conditions, this limit due to PHARAO systematic effects can be reached in terms of statistical uncertainty, when considering realistic noise levels for the space clock and clock comparison system.

\subsection{ACES payload\label{sb-payload}}

ACES will realize an ultra-stable time scale using two atomic clocks: a
hydrogen maser (SHM) and a cold atom Cesium clock (PHARAO). It contains
also a GNSS (Global Navigation Satellite System) receiver for orbit determination and a Microwave Link (MWL) module for two way time and frequency transfer to ground terminals. Analysing raw data from this MWL retrieves the scientific products, among which the desynchronization
between ground clocks and and the on-board time scale \cite{Duchayne2009, Meynadier2018}. 

Onboard ACES, the Frequency Comparison and Distribution Package (FCDP) allows the comparison of the SHM and PHARAO clock signal. It realizes a
short-term servo loop that phase locks PHARAO's local oscillator on SHM's
clock signal, and a long-term servo loop that frequency stabilizes the SHM
local oscillator on PHARAO's clock signal. FCDP combines both servo
loops in order to generate the ACES clock signal, which inherits the
short term stability of SHM and the long term stability and accuracy
of PHARAO. For averaging times exceeding 10 days, the stability will be limited by the PHARAO systematic uncertainty of $10^{-16}$. FCDP provides this ACES clock signal to the MWL for
comparison to ground clock signals.

\subsection{ACES data}

In the ACES mission, a given ground clock in a laboratory equipped
with a MWL ground terminal will be compared to PHARAO when ISS is in
sight. The time transfer software for extracting the clock
desynchronization from MWL raw data has been presented in
\cite{Meynadier2018}.

\begin{figure}[h!]
\begin{center}
\subfigure[]{\includegraphics[scale=0.7]{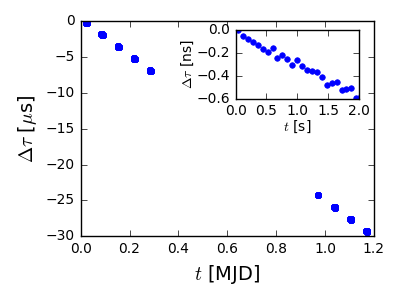}}
\subfigure[]{\includegraphics[scale=0.7]{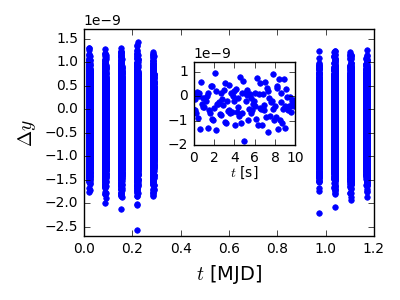}}
\end{center}
\caption{Simulated ACES data over one day for a ACES-OPMT clock
  comparison: (a) desynchronization, (b) differential fractional
  frequency shift.  On each graph the inset shows a zoom over a few seconds of one
  pass.}
\label{fig-data}
\end{figure}

The altitude of ISS is kept between 330 and 435 km, with an
inclination of $51.6^{\circ}$ with respect to the Earth equatorial plane,
and an orbital period of about 90 minutes. For a given ground station,
the passes last in average about 300 seconds (up to 500 seconds), with
each day a series of typically 5-6 passes approximately every 90
minutes interrupted by longer out of view periods. The ground-PHARAO
clock desynchronizations are measured every 80~ms of the local GT time
scale, which will be considered here to be UTC. We call these ``phase
data'', from which relative frequency differences (``frequency data'')
are derived. Simulated phase and frequency data, for 9 successive
passes of PHARAO above a ground station located in Paris (OPMT), are
shown on Figure \ref{fig-data}. As can be seen, data have large gaps
covering about 97\% of the total duration.

Currently six fixed and two mobile MWL ground terminals are
planned. The fixed ones will be located in National Metrology
Institutes, equipped with high performance atomic clocks (\textit{e.g.} Cesium/Rubidium
fountains or optical clocks) with stabilities that outperform the ACES
time scale, leaving the on-board clock and the MWL as the dominating
noise contributors. For the tests presented here we considered 10
possible locations listed in Table \ref{tab-stations}. The MWL can
operate on up to four channels, such that up to 4 ground stations can
be compared to ACES simultaneously.


\Table{\label{tab-stations} List of ground stations considered here,
  that could be possibly equipped with a MWL ground terminal. The
  first column indicates the label used in this article and the third
  column indicates the city and country. Columns 4 to 6 represent the station positions in ITRF}
\br
\textbf{Label} & Laboratory &\textbf{Location} & Latitude ($^{\circ}$) & Longitude ($^{\circ}$) & Height (m)\\
\mr
OPMT  & OBSPM & Paris, FR  		& \hspace*{\fill} 48.8 & \hspace*{\fill}   2.3 &\hspace*{\fill}   124.2	\\
PTBB & PTB & Braunschweig, DE 	& \hspace*{\fill} 52.3 & \hspace*{\fill}  10.5 &\hspace*{\fill}   130.2	\\ 
HERS & NPL & Hailsham, UK   	& \hspace*{\fill} 50.9 & \hspace*{\fill}   0.3 &\hspace*{\fill}    76.5	\\ 
NISU & NIST & Boulder, US  		& \hspace*{\fill} 40.0 &\hspace*{\fill} -105.3 &\hspace*{\fill}  1648.5	\\ 
TABL  & JPL & Wrightwood, US 	& \hspace*{\fill} 34.4 &\hspace*{\fill} -117.7 &\hspace*{\fill}  2228.0	\\
MTKA & NICT & Mitaka, JP  		& \hspace*{\fill} 35.7 &\hspace*{\fill}  139.6 &\hspace*{\fill}  109.0	\\ 
IENG & INRIM & Torino, IT 		& \hspace*{\fill} 45.0 &\hspace*{\fill}    7.6 &\hspace*{\fill}  316.6	\\ 
GRAS & OCA & Caussols, FR 		& \hspace*{\fill} 43.8 &\hspace*{\fill}    6.9 &\hspace*{\fill}  1319.3	\\ 
TSKB & GSI & Tsukuba, JP  		& \hspace*{\fill} 36.1 &\hspace*{\fill}  140.1 &\hspace*{\fill}    67.3	\\ 
PERT & UWA & Perth, AU  		&\hspace*{\fill} -31.8 &\hspace*{\fill}  115.9 &\hspace*{\fill}  12.9	\\ 
\br
\end{tabular}
\end{indented}
\end{table}

Auxiliary data (\textit{e.g.} temperature, pressure, power levels, etc...) that
allows correcting for numerous systematic effects will be
available. The ones that are important in the context of this work are
the ISS orbit and attitude data obtained from GNSS (Global Navigation
Satellite System) receivers on the ISS and the ACES payload
itself. Using that data the orbit of the PHARAO space clock reference
point (the centre of its microwave cavity) can be retrieved.

\subsection{Specifications\label{sb-spec}}

\begin{figure}[h!]
\begin{center}
\subfigure[]{\includegraphics[scale=0.51]{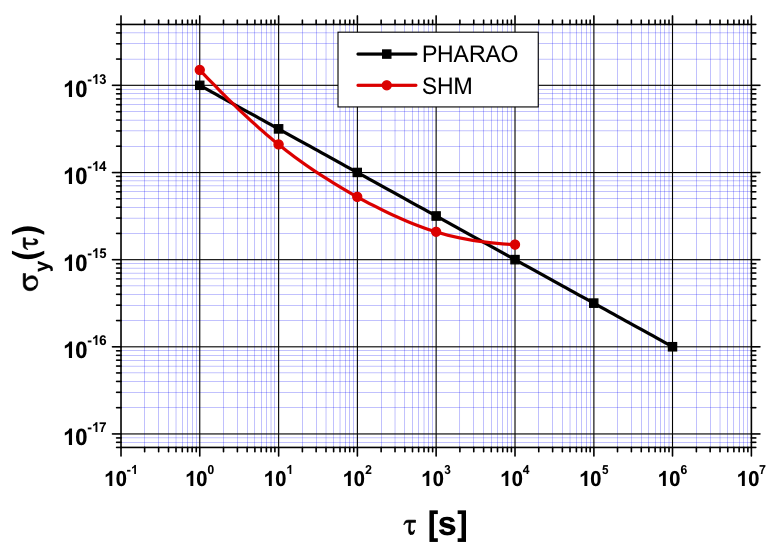}\label{fig-avar}}
\subfigure[]{\includegraphics[scale=0.51]{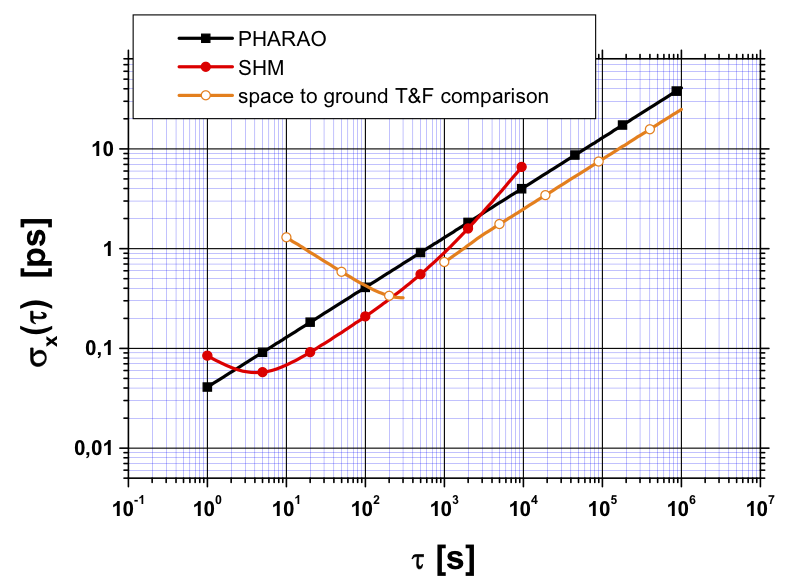}\label{fig-tdev}}
\end{center}
\vspace{-.5 cm}
\caption{(a) Allan deviation $\sigma_y(\tau)$ of SHM and PHARAO as a
  function of the averaging time $\tau$. (b)
  Time deviation $\sigma_x(\tau)$ of PHARAO, SHM and MWL.}
\label{fig-spec}
\end{figure}

The frequency stability specifications of the two ACES clocks are shown on
Figure \ref{fig-avar} \cite{TN}. For short term variations (below a few
$10^3$~s), SHM is more stable than PHARAO, whereas PHARAO
is the more stable clock for long averaging times. PHARAO stability is characterized by white frequency noise with a stability goal
of
$\sigma_y(\tau)=1\times10^{-13}/\sqrt{\tau}$ where $\tau$ is the averaging time
in seconds. The target accuracy is $1\times10^{-16}$ and is reached
after $\sim$ 10 days averaging. 

The time stability specifications $\sigma_x(\tau)$ are shown on Figure
\ref{fig-tdev}  \cite{TN}.
Over one pass ($\sim$ 300~s) the MWL noise is
white phase noise and its time deviation should average down to
$0.3$~ps at 300~s. For longer averaging times the dominant noise comes
from PHARAO.

\subsection{Overview of the simulation and analysis software} \label{sb-arch}
We have developed a software that simulates ACES desynchronization data as will be obtained from raw MWL data using the methods described in \cite{Meynadier2018}. It uses as basic input ISS orbitography files, ground station positions, a geopotential model, and PHARAO and MWL noise models. The software also allows calculating the model that needs to be adjusted to the data in order to search for a putative violation of the gravitational redshift, and carries out such an adjustment providing an estimate of the redshift violation parameter $\alpha$ and its uncertainty. 

It is written in Python3 and has approximately 3000~lines of code. It
can be run in three modes:
\begin{itemize}
\item[$-$] \emph{simulation}: simulates realistic data;
\item[$-$] \emph{analysis}: reads experimental or simulated data, and constructs the
  vector of observables and the model matrix that needs to be adjusted to it;
\item[$-$]  \emph{adjustment}: adjusts the model matrix to the vector of observables and evaluates parameter values and uncertainties.
\end{itemize}

\begin{figure}[h!]
\begin{center}
\subfigure[]{\includegraphics[scale=0.41]{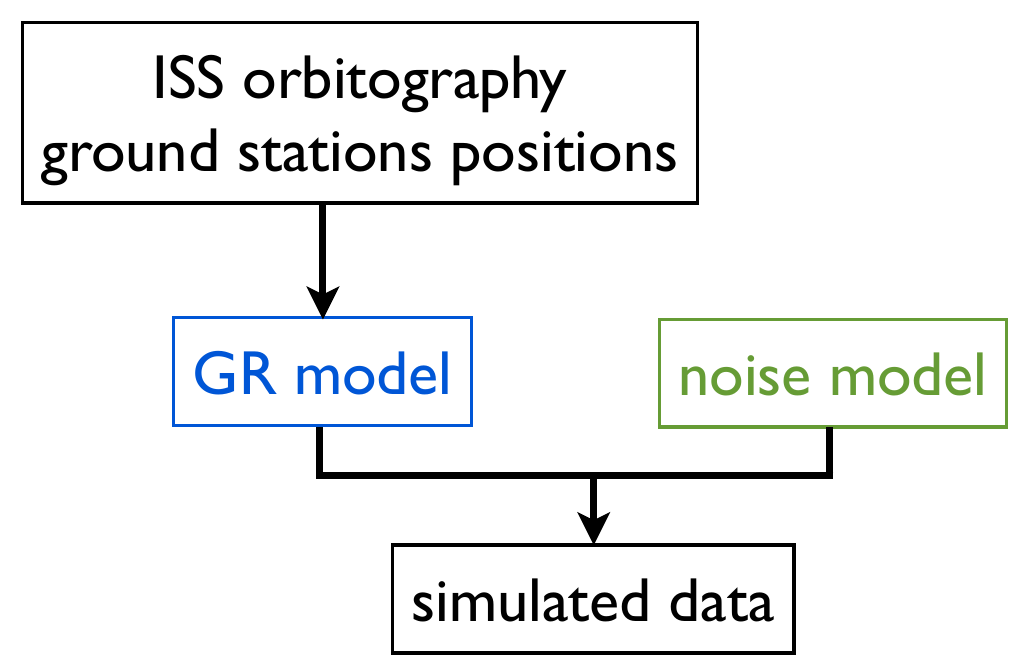}\label{fig-simu}}
\subfigure[]{\includegraphics[scale=0.41]{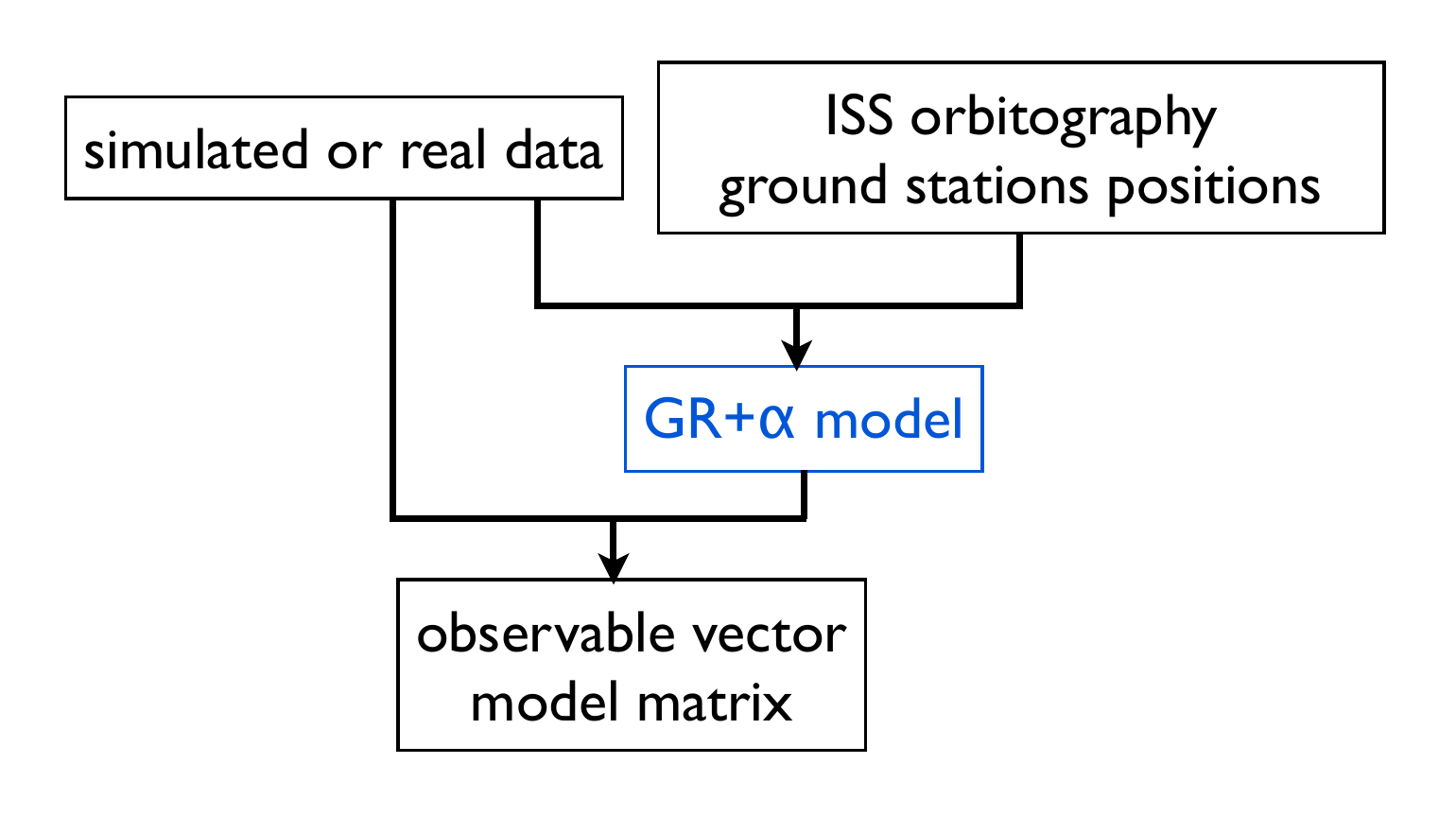}\label{fig-ana}}
\subfigure[]{\includegraphics[scale=0.41]{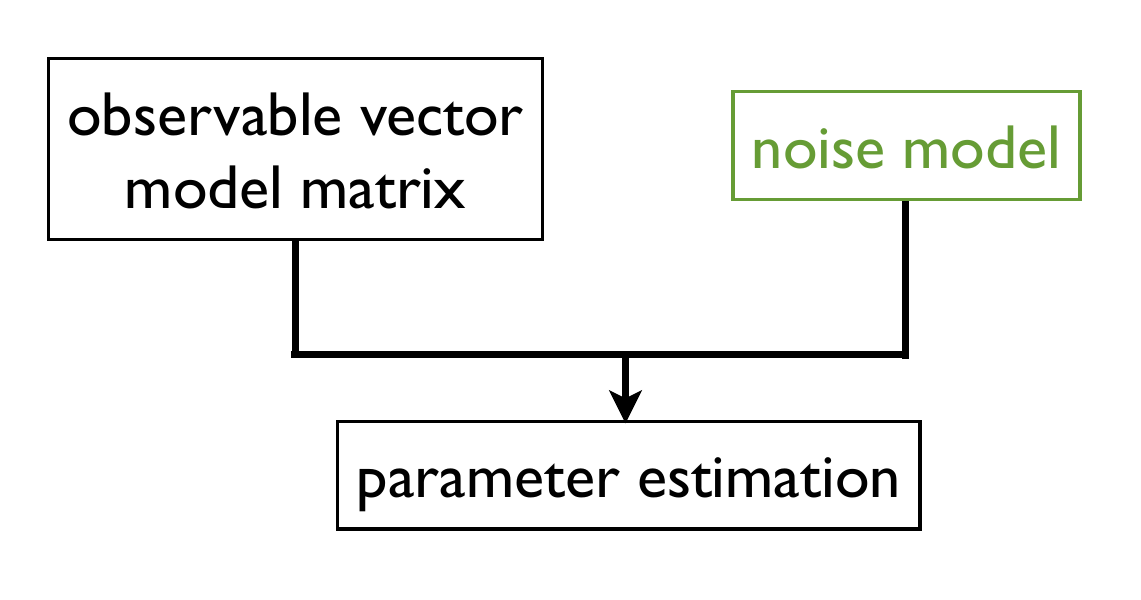}\label{fig-adj}}
\end{center}
\vspace{-.5 cm}
\caption{Schematic diagrams of the three operation modes of the software: (a) simulation,
  (b) analysis, (c) adjustment.}
\label{fig-diagram}
\end{figure}

\noindent It is modular and makes use of common Python classes. As
pictured on Fig. \ref{fig-diagram}, the simulation
and analysis modes have in common the construction of the GR model, and
the simulation and adjustment modes have in common the generation of
noise. Note that to estimate the effect of an orbitography error
(Sec. \ref{orbito}), the
analysis mode can be provided with an orbitography file that is different from the one used in the simulation mode, thus emulating imperfect knowledge of the orbit.

\section{Data simulation} \label{simulation}

The MWL data processing presented in \cite{Meynadier2018} will provide for each clock a file with 
coordinate time tags (in UTC) and
corresponding ground clock/PHARAO desynchronization, every 80~ms
during each pass, with a new file for each day.

However since no such real data are available yet, we simulate desynchronization data
as will be provided by ACES MWL data analysis \cite{Meynadier2018}, as well as
corresponding frequency data (discrete time derivative of the desynchronization data). For this we use
the GR model of Equation (\ref{eq-deltatau}), expressed in GCRS. This requires the knowledge of PHARAO and ground station
orbitography in GCRS and the choice of a
geopotential model (Subsection \ref{sb-pot}). The latter is given in a reference system rotating with the Earth, like the International Terrestrial Reference Frame (ITRF), see Subsection \ref{sb-coord}. We generate realistic time distribution of the
data regarding repetition rate and pass distribution (Subsection
\ref{sb-simu}). We also simulate different types of noise that affect the data (Subsection \ref{sb-noise}).

The input information for the simulation is the following:
\begin{itemize}
\item[$-$] fixed ground station position coordinates in ITRF2014 \cite{ITRF2014};
\item[$-$] a real ISS orbitography file in SP3 format \cite{SP3}, with the spatial coordinates given in ITRF, and GPS time;
\item[$-$] the Earth Geopotential Model 2008 (EGM2008) \cite{EGM2008};
\item[$-$] Earth Orientation Parameters (EOP) provided by IERS \cite{IERS2019} for frame
  transformations;
\item[$-$] instrumental noise levels.
\end{itemize}

For the simulation we used a 12-day ISS orbitography file provided by O. Montenbruck.

\subsection{Simulation and realistic data distribution\label{sb-simu}}
We first simulate for each clock independently (ground clocks and PHARAO) the fractional
frequency shifts from gravitational redshift and second order Doppler
effect, whose sum, from Eqs. (\ref{eq-dtaudt}) and (\ref{eq-ydef}), is:
\begin{equation}
y(t)=-\frac{
  U(t)}{c^2}-\frac{{\bf v}(t)^2}{2c^2}+  \Or(c^{-4}).
\label{eq-y}
\end{equation}

\noindent We then calculate for each ground clock the PHARAO-ground clock difference for
each term, and
integrate them in order to have their contributions to
desynchronization. This is done at every time
of the orbitography file (time step 30~s for the files we
used). Data are then cut in ``passes'' in
order to keep them, for a given ground station, only when ISS is in
line of sight, with a minimum elevation angle of $5^{\circ}$. Each column
(gravitational redshift differences and second order
doppler shift differences, in phase and frequency) is then interpolated in order
to have data every 80~ms. The sum of the two terms then gives the 
the overall differential frequency shift and desynchronization.

The noise on the PHARAO-ground clock difference is simulated every
80~ms and then cut with the same procedure. We simulate a single
noise file for PHARAO, added to individual noise files for each ground
station representing the MWL noise on their channel. 

The main outputs of the simulation are time series of noisy fractional frequency
differences $\Delta y(t_i)$ and desynchronizations $\Delta\tau(t_i)$, for each ground clock-PHARAO couple, dated 
in UTC at 80~ms sampling. 

For this simulation (and for the analysis), we use directly the
orbitography file of the center of mass of ISS. For real data we will need to calculate the
orbitography of the PHARAO clock, which requires the knowledge of
the ISS attitude.

The model part of this software (without noise) has been
tested against an independent software by Anja Schlicht and
Stefan Marz at the Technical University of Munich (TUM). We checked that, in frequency, the potential and Doppler
terms agree. This was done up to order 50 in potential for the ISS and 5 for
the OPMT ground station. The space-ground gravitational redshift
agrees within $2\times10^{-24}$ and the Doppler
shift agrees within $1.5\times10^{-18}$. 


\subsection{Geopotential\label{sb-pot}}

For the geopotential we use the model ``Earth Geopotential Model
2008'' (EGM 2008) \cite{EGM2008}, which gives the spherical harmonic coefficients of the static Earth potential at a given position in the reference system  World Geodetic System 84 (WGS84) \cite{WGS84}. The difference between
recent ITRF realizations and WGS84 being within 10~cm, we use directly
the ITRF
positions of ground clocks and PHARAO in order to calculate the
geopotential at their position. 

\begin{figure}[h!]
\begin{center}
\subfigure[]{\includegraphics[scale=0.7]{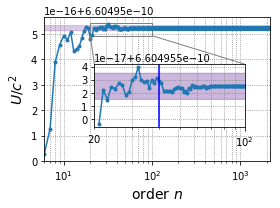}\label{fig-potISS}}
\subfigure[]{\includegraphics[scale=0.7]{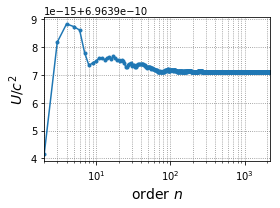}\label{fig-potOPMT}}
\subfigure[]{\includegraphics[scale=0.7]{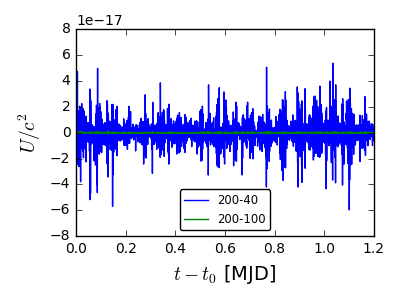}\label{fig-potconv}
}\end{center}
\vspace{-.5 cm}
\caption{Gravitational redshift $U/c^2$ depending on the truncation
  $n$ in the spherical harmonic expansion of the geopotential, for
  ISS at the first point of our orbitography file (a) and for OPMT ground station (b). On (a), the shadowed region
  is within $\pm10^{-17}$ from the convergence value. On the main
  graph the order ranges from 6 to 2190. The inset shows
  a zoom on the convergence, from order 20 to
  200. The vertical blue line marks order 40 after which the shift sits
well within in the convergence zone. (c): For the POD
  orbitography file, difference between the redshift calculated at
  order 200 and 40, and order 200 and 100.}
\label{fig-geopot}
\end{figure}

The
gravitational redshift $U/c^2$ for the OPMT ground station and for the
ISS, depending on the truncation order of the development, is shown
respectively on Figures \ref{fig-potISS} and \ref{fig-potOPMT}. The
OPMT station is considered to have a fixed position with respect to ITRF, thus its potential has to be calculated only once for the simulation and analysis. Therefore we calculate the OPMT potential at the maximum order 2190. Whereas for
ISS it has to be calculated at every point of the orbitography file
(which with a spacing of 30~s over 12 days amounts to ~$10^6$
points) which takes several hours when using all orders. 

To reduce the computational burden we
checked whether the calculation can be safely truncated at a lower
order without biasing the redshift analysis. The inset of
Fig. \ref{fig-potISS} shows that above order 40, the redshift
stays well within $10^{-17}$ from the convergence value. Above 200, at
this scale the variations of the redshift are not visible
any more. Thus a safe choice for a realistic simulation can be to
simulate and analyse at order 200. We checked on
Fig. \ref{fig-potconv} that for all points of the orbitography used,
the difference between the redshift at order 40 and 200 stays 
within a few parts in $10^{17}$, and the comparison between order 100 and 200
confirms that for the desired precision the potential has indeed
converged. As a final check, we run the full software,
simulating with order 200 and analyzing with order 10 or higher: we obtain the same (non-significant) $\alpha$ value and uncertainty irrespective of the order used.

As a consequence all tests presented below have been realized with order 200 for the ISS both for simulation and analysis. This is both self-consistent, and would retrieve safe values when analysing real data.

\subsection{Coordinate transformations\label{sb-coord}}

As seen in this section, the reference systems involved are GCRS,
ITRF, and WGS84; as explained in Subsection \ref{sb-pot}, WGS84 is
approximated here to ITRF.  The time scales
involved are UTC, TCG (geocentric coordinate time, associated to GCRS), and GPS time. GPS time is a
continuous atomic time scale, constantly late versus TAI by 19~s. 

One space-time coordinate transformation is required: for the Doppler effect term in Eq. (\ref{eq-y}), ground stations
and ISS ITRF velocities have to be transformed into GCRS velocities.

In the integral of Eq. (\ref{eq-deltatau}) we need the coordinate time
associated to the GCRS frame \textit{i.e.} the TCG time scale. Thus
the GPS time of the ISS orbitography file has to be converted into
TCG. 

For all coordinate transformations we use the coordinate transformation package developed for the MWL
simulation and analysis softwares, based on the SOFA packages from the IAU (for details see \cite{Meynadier2018}).

\subsection{Noise model and simulation\label{sb-noise}}

We take into account two noise contributors: the PHARAO clock and the MWL, thereby neglecting the
noise from ground clocks which is not expected to be limiting. We conservatively take
the ACES clock noise to be the one of PHARAO at all times although in the 
short term it will be a bit better thanks to SHM (see Subsection
\ref{sb-spec}). 

As seen in Subsection \ref{sb-spec}, the PHARAO noise is white frequency
noise. The MWL noise will be dominant at short times, for which it is
white phase noise. We approximate the MWL noise as a white
phase noise at all times, thereby neglecting other contributions that
only appear in the longer term and will remain negligible for our test, as the long term is dominated by PHARAO noise. 
For the noise level of PHARAO, we take the
mission specifications presented in Subsection \ref{sb-spec}. As the
MWL is, at the present status, not nominally working yet, and different specifications can be found in the literature, we take a slightly more conservative noise level of about 0.4 ps at 300
seconds (instead of 0.3 ps in Subsection \ref{sb-spec}). 


The white frequency noise from PHARAO is simulated in frequency and
then integrated to get its contribution to phase data. The white phase
noise from the MWL is simulated in phase and derived in order to get its
contribution to frequency data. 

\begin{figure}[h!]
\begin{center}
\subfigure[]{\includegraphics[scale=0.75]{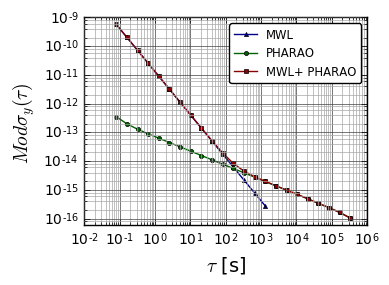}\label{fig-adevsim}}
\subfigure[]{\includegraphics[scale=0.75]{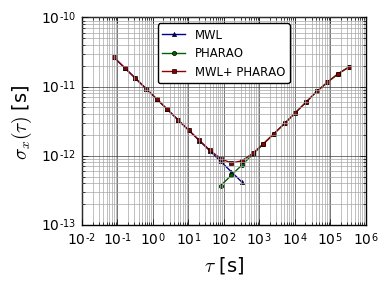}\label{fig-tdevsim}}
\end{center}
\vspace{-.5 cm}
\caption{(a): Modified Allan deviation of the noise on relative
  frequency difference. (b): Time deviation of the noise on
  desynchronization. 
  }
\label{fig-devsim}
\end{figure}

On frequency data, we have the sum of a white noise from PHARAO (PSD in $f^0$), and
violet noise from the MWL (PSD in $f^2$). Its modified Allan deviation is shown
on Figure \ref{fig-adevsim}. On phase data, we have the sum of a white
noise from the MWL, and random walk noise from PHARAO, with respective
Power Spectral Densities (PSD) in $f^0$ and $f^{-2}$. Its time
deviation is shown on Figure \ref{fig-tdevsim}.

As can be seen on this figure, the
noise from PHARAO dominates after 300~s (approximately one
pass). The dominant long term noise is thus white noise for frequency
data, and random walk noise for phase data.

When considering ground-space clock comparison data for several ground
clocks during a common time span, the PHARAO clock is in common mode,
though not necessarily sampled at the same times. On the other hand
since there are up to 4 MWL channels with each a given noise, we consider
the MWL noise to be independent for each clock pair. When simulating
noise for a set of ground clock-PHARAO comparisons on a given time
span, we thus simulate a single PHARAO noise, and one MWL noise per
ground-PHARAO pair.

\section{Data analysis: methods and assessment} \label{modeling}
We present here the key elements of our modeling method. We do not
present in detail its software implementation, which as
explained in Subsection \ref{sb-arch}, shares common routines with the
simulation part presented in Section \ref{simulation}.

\subsection{Data types and models\label{sb-models}}
We performed the analysis on desynchronisation data $\Delta\tau(t_i)$ and on
relative frequency difference data $\Delta y(t_i)$, which is its time
derivative. We define our observable
as the data minus the GR model, allowing us to keep the maximum numerical
resolution with values closer to zero .

\begin{figure}[h!]
\begin{center}
\subfigure[]{\includegraphics[scale=0.7]{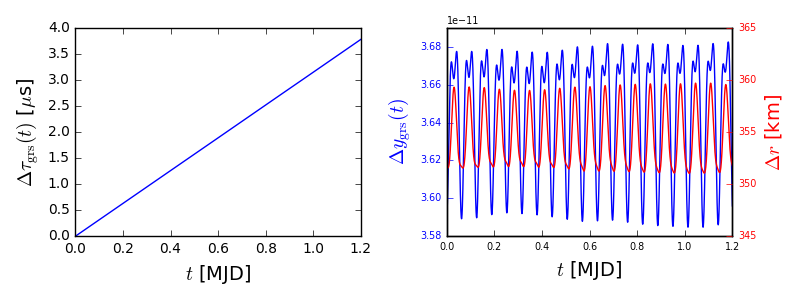}\label{fig-datapotm}}
\subfigure[]{\includegraphics[scale=0.7]{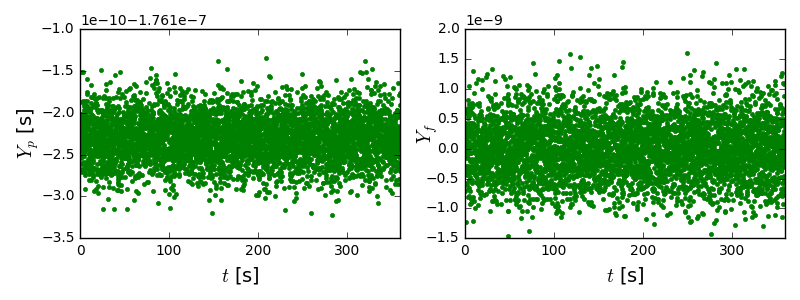}\label{fig-y}}
\end{center}
\caption{Modeling of the PHARAO-OPMT clock comparison. (a) Gravitational redshift model over one day. Left: for phase data. Right: for frequency data, with
  on the right axis the difference in norm of the ISS and OPMT
  position vectors. (b) Simulated phase (left) and frequency (right) observables during
the first pass.}
\end{figure}

At a given coordinate time $t_i$ and for a given
ground station, our phase observable is thus:
\begin{equation}
Y_p(t_i)=\Delta\tau(t_i) +\int_{t_0}^{t_i} (\frac{\Delta U
  (t')}{c^2}+\frac{\Delta {\bf v}(t')^2}{2c^2})dt'
\label{eq-obsp}
\end{equation}
\noindent where $t_0$ is the time of the first data point. Our frequency observable is:
\begin{equation}
Y_f(t_i)=\Delta y(t_i)+ \frac{\Delta U(t_i)}{c^2}+\frac{\Delta {\bf
    v}(t_i)^2}{2c^2}.
\label{eq-obsf}
\end{equation}
\noindent For the
gravitational potential calculation, we use the same spherical
harmonics orders than the
one used for the simulation (\textit{i.e.} 200 for  ISS and 2190 for ground
stations). The two observables are shown on Fig. \ref{fig-y}. The
standard deviation for phase data is $3\times 10^{-11}$~s which corresponds to
the Allan time deviation at 80~ms; for frequency data it is $5\times 10^{-10}$, close to
the modified Allan deviation at 80~ms which is $6\times 10^{-10}$ (see Fig. \ref{fig-devsim}).

For each observable, the model we want to adjust includes the gravitational
redshift term multiplied by
the violation parameter $\alpha$, and possible overall offsets. For a given
ground station-PHARAO pair, the model in phase has two parameters, the
initial clock desynchronization $\Delta\tau_0$ and the gravitational
redshift parameter $\alpha$:
\begin{equation}
Y_p^m(t)=\Delta\tau_0 - \alpha \int_{t_0}^t \frac{\Delta U (t')}{c^2}dt'.
\end{equation}
The model in frequency has only the parameter $\alpha$:
\begin{equation}
Y_f^m(t)=-\alpha \frac{\Delta U(t)}{c^2}.
\end{equation}
\noindent If GR and more precisely LPI is verified,
$\alpha$ should be zero. 

The model functions for the gravitational redshift term are shown on
Fig. \ref{fig-datapotm}: $\Delta\tau_{\mathrm{grs}}(t)=-\int_{t_0}^t
\frac{\Delta U (t')}{c^2}dt'$ for phase data and $\Delta
y_{\mathrm{grs}}(t)=- \frac{\Delta U(t)}{c^2}$ for frequency data. As can be seen,
the model in frequency, proportional to the potential, is mainly a
constant offset, with a small modulation coming from
the slight ellipticity of the ISS orbit, as visible from the
height difference between ISS and OPMT also plotted. The gravitational redshift
parameter $\alpha$ is thus mostly determined by the average of the
frequency obervable $Y_f(t_i)$. For phase, the model is
mainly a linear drift (with a small periodic modulation due to the
orbit ellipticity, not visible at this scale). The gravitational
redshift parameter $\alpha$ is thus in this case mostly determined by the slope
of the phase observable $Y_p(t_i)$.

When analyzing simultaneously the data from the different ground
stations, we can do the analysis independently for each clock, or
do a global adjustment. In the latter case we adjust a common
$\alpha$ coefficient and one time offset per ground station.

\subsection{Noise}
In the analysis, since the space-ground clock comparison data have large gaps, estimating all noise
levels and colors from the data themselves is difficult. Analyzing the
data time stability should allow to
retrieve the MWL time
deviation over a typical pass duration, where it is dominant. PHARAO's
white frequency noise level
will be independently estimated on ISS via its comparison to the SHM clock,
which is more stable at short times (see
Subsec. \ref{sb-spec}). PHARAO's frequency noise is characterized as white as long as not
systematics-dominated; systematics will be independently evaluated
by varying clock operation parameters and are not expected to be
dominant before ~10 days. Thus measuring the noise behavior at short
times allows to infer PHARAO's frequency stability during a
typical session time. 

Here we investigate how well we can
determine $\alpha$ under given noise assumptions, so we use the known
noise levels and colours that went into the simulation according to
the mission specifications, as described in Section
\ref{sb-noise}. One important point to notice for the analysis is that
both for phase and frequency data, the noise is not white at all averaging times (see fig. \ref{fig-devsim}).

\subsection{Statistical method}\label{sb-model}

As the observables $Y_{f,p}(t)$ in Eqs. (\ref{eq-obsp}), (\ref{eq-obsf}) depend linearly on the parameters,
we can use a linear least-squares estimator. Under matrix form, 
the general equation describing an observable $Y$ (length $N$) is:
\begin{equation}
Y=X\beta+\epsilon
\label{eq-syst}
\end{equation}
with $\beta$ the vector of parameters to be estimated (length $p$), $X$ the model
matrix (dimensions $N\times p$). The noise vector $\epsilon$ (length
$N$) is supposed gaussian (thus $E[\epsilon]=0$), and has a covariance
matrix $\Omega=E[\epsilon\epsilon^T]$. 

In our case, for one station we have in Eq. (\ref{eq-syst}) for
frequency data the following matrices:
\begin{equation}
Y=\begin{bmatrix}Y_f(t_1) \\
  Y_f(t_2)\\...\\Y_f(t_N)\end{bmatrix}\mbox{, }X=\begin{bmatrix}\Delta
  y_{\mathrm{grs}}(t_1)\\\Delta y_{\mathrm{grs}}(t_2)\\...\\\Delta
  y_{\mathrm{grs}}(t_N)\end{bmatrix}\mbox{, }\beta=\begin{bmatrix}\alpha\end{bmatrix},
\end{equation}
and for phase data:
\begin{equation}
Y=\begin{bmatrix}Y_p(t_1) \\
  Y_p(t_2)\\...\\Y_p(t_N)\end{bmatrix}\mbox{, }X=\begin{bmatrix}\Delta
  \tau_{\mathrm{grs}}(t_1)&1\\\Delta \tau_{\mathrm{grs}}(t_2)&1\\...\\\Delta
  \tau_{\mathrm{grs}}(t_N)&1\end{bmatrix}\mbox{, }\beta=\begin{bmatrix}\alpha\\\Delta\tau_0\end{bmatrix}.
\end{equation}
The number of data points $N$ for one station
 during the 12 days of our orbitography file is about $3\times10^5$.

The aim is to determine an estimator $\hat{\beta}$ of $\beta$, as well as the uncertainty and
correlations of its components, which are respectively the square root
of the diagonal and off-diagonal components of the variance-covariance
matrix defined by $V=E[(\hat{\beta}-\beta)
(\hat{\beta}-\beta)^T]$. The Ordinary Least Squares (OLS)
estimator is:
\begin{equation}
\hat{\beta}_{\mathrm{OLS}}=(X^TX)^{-1}Y.
\label{eq-bols}
\end{equation}
Its variance-covariance matrix is 
\begin{equation}
V_{\mathrm{OLS}}=\sigma^2(X^TX)^{-1}
\label{eq-cols}
\end{equation}
 if the
noise is uncorrelated (diagonal variance-covariance
matrix $\Omega=\sigma^2{\bf{1}}$).

The noise of our data is a correlated gaussian noise: its average is zero but its covariance matrix is not diagonal. In this case, the
Ordinary Least Squares (OLS) estimator in Eq. (\ref{eq-bols}) is still
unbiased\footnote{The additional assumption we fulfill is that our
  model $X$ is not stochastic.}: $E[\hat{\beta}_{\mathrm{OLS}}-\beta]=0$,
but the covariance formula in Eq. (\ref{eq-cols}) is not correct
any more.

We tested and compared two extensions of the least-squares method adapted to the case of
correlated noise: the Generalized Least Squares method (GLS), and Least Squares
Monte-Carlo (LSMC).


\subsubsection{Generalized Least Squares}

The Best Linear Unbiased Estimator (BLUE) for our data is the
GLS estimator. It reaches the Cramer-Rao lower bound,
which is the lowest possible variance attainable from our
data. The GLS method is equivalent to applying OLS to a linearly transformed version of the system
(\ref{eq-syst}) where the transformed noise has a diagonal covariance matrix.
The GLS estimator is:

\begin{equation}
\label{eq-GLSparam}
\hat{\beta}_{\mathrm{GLS}}=(X^T\Omega^{-1}X)^{-1}X^T\Omega^{-1}Y
\end{equation}

\noindent and its variance-covariance matrix is:
\begin{equation}
\label{eq-GLSv}
V_{\mathrm{GLS}}=(X^T\Omega^{-1}X)^{-1}
\end{equation}

\noindent where $\Omega$ is the covariance matrix of the noise.

The noise vectors for our data are series of the sum of two
integer-power noise components (white and violet for frequency data, white and random
walk for phase data) with large gaps. Each individual covariance
matrix as well as its inverse has a simple analytical form (given in
\ref{sec-rw} and \ref{sec-vn}) unlike the covariance matrix of the total noise:
the latter is the sum of the individual covariance matrices, and there is no
simple analytical expression of its inverse.

 The total covariance matrix of the
frequency noise is sparse, but not for phase data which due to the random walk
noise has all $N^2$ matrix elements non-zero. Numerical inversion
of such a
matrix is memory demanding and therefore limited
in data length to $\sim3\times10^4$ points which covers, for our data
time distribution, approximately one
day for one ground station. We will refer to this method in the
following as ``exact
numerical GLS''.

For longer time series, an approximated
version of this method can be implemented. Indeed, as seen on Figure
\ref{fig-devsim}, for phase data the random walk noise dominates after 300~s.  The white
phase noise contribution can thus be neglected on the long term, which
is expected to be dominant for determining $\alpha$ (dependent mainly
on the overall slope as seen in Subsection \ref{sb-models}). In
this approximation, we can use the simple
analytical form of the inverse of the covariance matrix of random walk
noise (\ref{sec-rw}). We will refer to this method in the
following as ``approximated analytical GLS'' (AGLS).

\subsubsection{Least Squares Monte-Carlo}\label{sb-LSMC}

In the LSMC method, we determine the parameter values from the OLS estimator
(Eq. (\ref{eq-bols})) since it is unbiased, and the uncertainty from $n_{\mathrm{MC}}$ Monte-Carlo simulations of our
noise. We simulate $n_{\mathrm{MC}}$ noise vectors $(\epsilon)_n$, and
for each we adjust our model to the noise
vector with the OLS parameter estimator:

\begin{equation}
\label{eq-LSMCparam}
(\hat{\beta})_n=(X^TX)^{-1}X^T(\epsilon)_n.
\end{equation}

\begin{figure}[h!]
\begin{center}
\subfigure[]{\includegraphics[scale=0.7]{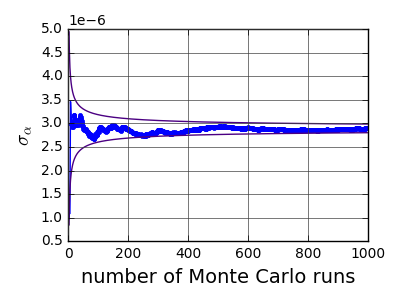}\label{fig-conv}}
\subfigure[]{\includegraphics[scale=0.7]{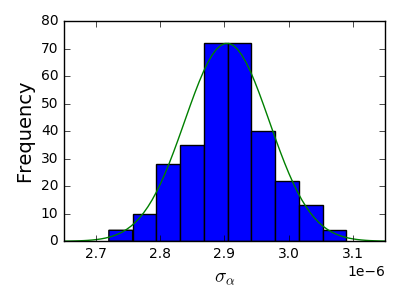}\label{fig-sigMC}}
\end{center}
\caption{ Convergence of the LSMC method. (a) Evolution of $\sigma_{\alpha}(n)$ for the OPMT station and phase
  data when adding a $n^{\mathrm{th}}$ Monte-Carlo run (blue line);
  the purple lines show the observed dispersion trend $\sigma_{\alpha}(N)(1\pm\frac{1}{\sqrt{n}})$. (b)
Histogram of $\sigma_{\alpha}$ results obtained from 300
  realizations of 1000 Monte-Carlo runs. The green line is a gaussian $Ae^{\frac{(x-\mu)^2}{2\sigma^2}}$
  with: $\mu=2.90\times10^{-6}$ and $\sigma=0.07\times10^{-6}$
  respectively the average and standard deviation of
the histogram.}
\end{figure}

\noindent For the $i^{\mathrm{th}}$ parameter, the uncertainty is estimated as the
standard deviation of the distribution of $(\hat{\beta}_i)_n$. For
phase data, we
estimate the covariance
between parameters $i$ and $j$ using the
$p\times n_{\mathrm{MC}}$ matrix $M$ of the Monte Carlo results:
$\mathrm{Cov}[\hat{\beta}_i,\hat{\beta}_j]=E[\hat{\beta}_i\hat{\beta}_j]=\frac{1}{N}(M^TM)_{i,j}$
with $M_{i,j}=(\hat{\beta}_i)_j$.
We investigate empirically the convergence of the Monte-Carlo method by plotting
 the evolution of the estimated uncertainty on the $\alpha$
parameter with the OPMT station for phase data, as the number of
Monte-Carlo runs increases (Figure
\ref{fig-conv}). We observe a dispersion scaling as $1/\sqrt{n}$
which at $n_{\mathrm{MC}}=10^3$ should reach $3\%$. We check this dispersion by
repeating 300 times the estimation
of $\sigma_{\alpha}$ by $n_{\mathrm{MC}}=10^3$
Monte-Carlo runs. The histogram of the 300 $\sigma_{\alpha}$ values is shown on Figure
\ref{fig-sigMC}. The standard deviation of the results is
$0.07\times10^{-6}$, \textit{i.e.} $2.3\%$ of their average
$2.90\times10^{-6}$, in accordance with the $1/\sqrt{n}$
convergence observed on Fig. \ref{fig-conv}. This relative spread being low, we therefore choose
$n_{\mathrm{MC}}=10^3$ for the LSMC method.

\subsection{Comparison of GLS and LSMC uncertainty}\label{sb-compGLS}
We compared the results of the LSMC method, for one
ground station (OPMT), both for phase and frequency data, with the two
versions of the GLS method presented in the Subsec. \ref{sb-model}:
\begin{itemize}
\item[$-$] ``exact numerical GLS'' over  $3\times 10^4$ points ($\sim$1 day);
\item[$-$] ``approximated analytical GLS'' (AGLS) over the full duration of our simulated data (12 days).
\end{itemize}

\noindent Here we present only the results for phase
data. Indeed, as will be shown in Section \ref{redshift}, only phase
data allow us to reach the targeted redshift test uncertainty. The results are shown in Table \ref{tab-GLS}.

The uncertainties obtained by the LSMC method for the redshift test
parameter $\alpha$ are very close (within a few \%) but higher than
the GLS methods, as expected given that GLS is the optimum
estimator. We note slight fluctuations of the LSMC results as a
function of the generated noise occurrences (see Section
\ref{sb-LSMC}). The large discrepancy between the uncertainties of the
initial offset, $\sigma_{\Delta\tau_0}$, when using LSMC or AGLS over
12 days are probably due to the large correlation coefficients for the
particular LSMC noise occurrences versus the practically zero
correlation coefficient for the analytical AGLS. As we are not
primarily interested in the initial phase offset we did not
investigate that question further.

In the following, we choose our parameter value estimator to be the OLS and
estimate our uncertainty with the LSMC
method. According to the test presented above, we know that this estimator
is conservative and close to optimal for the uncertainty in $\alpha$. There would be no big advantage in our case in using the
GLS, which has the drawback of being only an approximation for longer
times series, and requires to switch between exact and approximate
versions depending on the length of analyzed data since the pure
random walk noise hypothesis is not valid at short times.


\Table{\label{tab-GLS} Estimated uncertainty
($\sigma_{\hat{\beta}_i}$) and correlation coefficient
(Cor$[\hat{\beta}_i,\hat{\beta}_j]\equiv
\mathrm{Cov}[\hat{\beta}_i,\hat{\beta}_j]/(\sigma_{\hat{\beta}_i}\sigma_{\hat{\beta}_j})$)
obtained by the LSMC and exact numerical GLS methods over 1 day, and
LSMC and approximate analytical GLS (AGLS) over 12 days, for phase data with the
OPMT ground station.}
\br
&LSMC (1d)&GLS (1d) & LSMC (12d) & AGLS (12d)\\
\mr
$\sigma_{\alpha}$&$1.0\times10^{-5}$&$9.6\times10^{-6}$&$2.9\times10^{-6}$&$2.7\times10^{-6}$\\
$\sigma_{\Delta\tau_0}$ (s)&$7.2\times 10^{-12}$&$8.4\times 10^{-13}$&$3.4\times 10^{-11}$&$2.8\times 10^{-14}$\\
\br
Cor$[\alpha,\Delta\tau_0]$&-0.05&-0.03&-0.19&$-5.8\times 10^{-8}$\\
\br
\end{tabular}
\end{indented}
\end{table}

\section{Gravitational redshift test} \label{redshift}
In this section we use the simulated data (Sec. \ref{simulation}) and the
chosen analysis method (Sec. \ref{modeling}) to assess the precision
expected on the gravitational redshift. We present results of the
analysis on
phase and frequency data and show that they are not equivalent. We
investigate how the result in phase, which is much better, scales when
the data duration increases. We also
test the impact of using all ground stations instead of
one. 

\subsection{Analysis of frequency and phase data}
We analyze the simulated frequency and phase data, for one station
(OPMT), during 12 days (68 passes of ISS above OPMT),
with the LSMC method. The results are presented in the
upper part of Table \ref{tab-OPMT}. For phase data, the
uncertainty on $\alpha$ is $2.8\times10^{-6}$, within a factor 1.5 from the mission
target $2\times 10^{-6}$. For frequency data, the
uncertainty $\sim 5\times10^{-4}$ falls short by two orders of magnitude.

In order to understand this difference, we realized the same test but
with a different time distribution of the data. In the middle part of Table
\ref{tab-OPMT}, we present the results obtained for continuous data,
\textit{i.e.} as if PHARAO and OPMT clocks where in line of sight during the
full 12 days span. For phase data, the
uncertainty is barely modified, whereas in frequency the result is
much better than for gapped data and becomes comparable to the one in phase.
Thus for the gravitational redshift test, gaps play a major role for frequency data
but not for phase data. We can interpret this as follows. As seen in
Subsection \ref{sb-models}, in phase data
we estimate the slope of desynchronization, for which the uncertainty is determined mainly by the difference between final and initial
values and times. For frequency data we estimate the data average, whose
uncertainty is limited by the overall number of points. The uncertainty in
frequency is thus
equivalent to phase data if all data are present, but
it is degraded when removing points (gaps).

To verify this hypothesis, we realized a third test, where we keep
only the first and last pass. The results are given in the lower part
of Table \ref{tab-OPMT}. The uncertainty for phase data is not significantly changed, which
supports our interpretation. In frequency, due to the lower number of
points, the uncertainty is further decreased
compared to the realistic distribution case.

In the following, we will only analyze phase data, since frequency
data fall short of the mission target uncertainty by several orders of magnitude.

\Table{\label{tab-OPMT}Gravitational redshift test results for one station (OPMT)
  over 12 days, for
  frequency and phase data. We used three data time distribution:
  realistic, continuous, and only first and last pass, with overall
  number of points respectively on the order of $3\times10^5$, $10^7$
  and $9\times10^3$.}
\br
Data distribution&Parameter&Frequency data&Phase data\\
\ns\mr
realistic&$\alpha$&$(0.3\pm4.9)\times10^{-4}$&$(0.9\pm2.8)\times10^{-6}$\\
&$\Delta\tau_0$~(s)&&$-1.8\times10^{-7}\pm3.3\times 10^{-11}$\\
&\crule{3}\\
&Cor$[\alpha,\Delta\tau_0]$&&-0.24\\
\mr
continuous &$\alpha$&$(2.9\pm3.1)\times10^{-6}$&$(0.9\pm2.9)\times10^{-6}$\\
&$\Delta\tau_0$~(s)&&$-8.8\times10^{-9}\pm3.7\times 10^{-11}$\\
&\crule{3}\\
&Cor$[\alpha,\Delta\tau_0]$&&-0.28\\
\mr
first and last pass&$\alpha$&$(-1.2\pm2.8)\times10^{-3}$&$(1.5\pm2.6)\times10^{-6}$\\
&$\Delta\tau_0$~(s)&&$-1.8\times10^{-7}\pm1.2\times 10^{-12}$\\
&\crule{3}\\
&Cor$[\alpha,\Delta\tau_0]$&&-0.01\\
\br
\end{tabular}
\end{indented}
\end{table}

\subsection{Scaling with data duration}

\begin{figure}[h!]
\begin{center}
\includegraphics[scale=0.7]{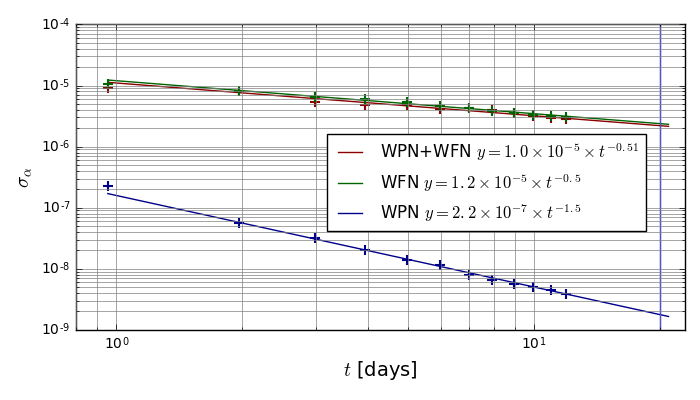}
\end{center}
\vspace{-.5 cm}
\caption{Scaling of the gravitational redshift test uncertainty in phase versus the
  length of data analyzed. The 12 days data set is truncated to
  shorter durations and analyzed. The results are plotted for
  individual noises and their sum, as a function of the time interval
  between the last and the first data point. Uncertainties on $\alpha$ are shown
  as a cross, and their power-law adjustments as a line. The
  adjusted formulae are displayed in the legend (with $t$ in
  days). The vertical blue line is set at 20 days.}
\label{fig-scaling}
\end{figure}

The uncertainty reached is 1.5 times higher than the mission target when we use a single 12 day data set (limited by the orbitography file duration). In reality several data sets in the 10-20 days range will be available during the 18 months to 3 years mission lifetime. We thus
investigate how our result will extend for longer data sets. For this, we
estimate the scaling with time of the uncertainty on $\alpha$, by
truncating the simulated OPMT data to shorter durations. We stop the data after
$1\leqslant t \leqslant 12$ days, and plot the obtained uncertainty on $\alpha$ versus the
duration $t$ between the last and first data points. The result is shown
on Figure \ref{fig-scaling}. Fitting for a power-law decay, we obtain
the following scaling (with $t$ in days): 
\begin{equation}
\label{eq-scaling}
\sigma_{\alpha}(t)=1.0\times10^{-5}\times
t^{-0.51}.
\end{equation}

In order to understand this scaling, we realized the same
test but considering separately the contribution from each noise:
white frequency noise (WFN) from PHARAO, and white phase noise (WPN)
from the MWL. The uncertainties and power-law adjustments are also
shown on Figure \ref{fig-scaling}. The WFN result is very
close to the result for both noises and appears to be the limiting noise;
this is expected since for data duration above 300~s it is the
dominant noise in terms of time deviation as seen in Subsection \ref{sb-noise}. The
scaling, close to $t^{-1/2}$, can be easily understood. Indeed for
phase data the WFN
appears as a random walk noise, whose standard deviation scales
with $t^{1/2}$; on the other hand, the slope estimation (fit of
$\alpha$ for phase data as seen in Subsection \ref{sb-models}) scales
with the duration $1/t$. This leads to the overall observed $t^{-1/2}$ scaling
on the slope estimated in the presence of WFN.

From the scaling in Eq. (\ref{eq-scaling}), for a 20 days duration one expects
a statistical uncertainty on $\alpha$ of $2.2\times10^{-6}$, very close to the
mission target. On the other hand combining results from independent
sessions will also improve the uncertainty of our test; the
uncertainty goal could \textit{e.g.} also be retrieved from 3 independent sessions of 12 days.


\subsection{Scaling with the station number}

\begin{figure}[h!]
\begin{center}
\includegraphics[scale=0.7]{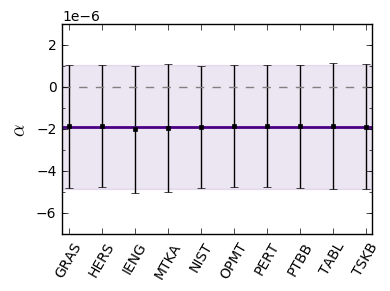}
\end{center}
\vspace{-.5 cm}
\caption{Values and uncertainty of $\alpha$
  for each individual station (black symbols and error bars), and
  from a global fit (violet line and shadowed region). The dotted
line is placed at 0 to check the (non-)significance of the values.}
\label{fig-Nsta}
\end{figure}

We compared the result obtained from each ground station, and from a
combined fit of all ground stations. The results are shown on Fig.
\ref{fig-Nsta}. 

All values of $\alpha$ are non-significant. When comparing individual results as well as the global fit results, all uncertainties are the same within
3\%, which is consistent with the intrinsic dispersion of the
$\sigma_{\alpha}$ estimate due to the LSMC method of $2.3\%$ (Subsection
\ref{sb-LSMC}). Note that the individual results (and the global fit) are not independent. They are all dominated by the same common noise due to PHARAO. This explains the very small dispersion of the points in Figure \ref{fig-Nsta}, when compared to the error bars.

The combined result shows no improvement with respect to single
stations. This is expected, since we have seen in Table
\ref{tab-OPMT} that no statistical gain was obtained for phase data when removing
all gaps from our data. Note that nonetheless, using data from different ground
stations will be useful and necessary to ensure robustness of our test when
assessing the systematics, and are also required for other science objectives of ACES \textit{e.g.} in time/frequency metrology.

\section{Estimation of the orbitography uncertainty requirement} \label{orbito}

Since the gravitational redshift depends on the ISS position,
imperfect knowledge of the ISS orbitography can lead to a bias in our
estimation of the deviation parameter $\alpha$. 

A naive estimate would be that since the gravitational redshift scales
with the inverse of distance ($U\sim GM/r$ with $G$ the gravitational
constant and $M$ the Earth's mass), the same relative uncertainty is required
on the knowledge of the distance between clocks as on $\alpha$. This would require an uncertainty of 2 ppm on
the ISS
range. The altitude being about 400~km, the orbit uncertainty
would have to be $\sim$1~m. 

This question was first treated
in \cite{Duchayne2009}, analytically, and numerically using two orbit error
models. Both the errors on the time transfer and on the gravitational
redshift were investigated. It was shown
that the requirement is less
stringent than expected ($\sim$10~m) since a radial position orbit error is associated
with a velocity error, and for the clock relativistic frequency shift, the errors from
the second order Doppler effect and the gravitational redshift
partially cancel. Once the MWL time transfer software was
developed, the impact on time transfer was addressed numerically with
the MWL software itself,
with several levels of uncertainty of a real ISS orbitography file. It showed
that below 1~km orbit error the impact is below specifications \cite{Meynadier2018},
confirming the estimates in \cite{Duchayne2009} for the time transfer part.

\begin{figure}[h!]
\begin{center}
\subfigure[]{\includegraphics[scale=0.7]{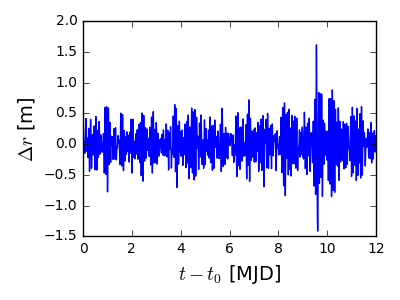}\label{fig-difforbito}}
\subfigure[]{\includegraphics[scale=0.72]{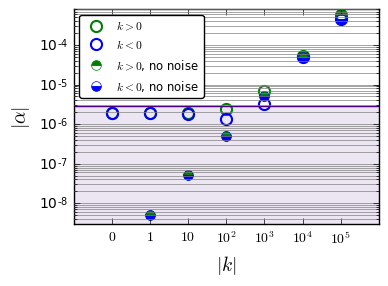}\label{fig-alphaorbito}}
\subfigure[]{\includegraphics[scale=0.72]{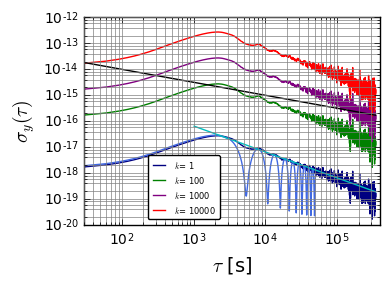}\label{fig-adevorberr}}

\end{center}
\vspace{-.5 cm}
\caption{(a): Position vector norm difference $\Delta r= r_{\mathrm{OD}}-r_{\mathrm{POD}}$ between the POD and OD
  ISS orbitography files. (b): Absolute values of $\alpha$ estimated when using
  the different orbitography files for the
  adjustment model, \textit{i.e.} POD, OD, or $k-$degraded versions of the OD
  file, with $k>0$ (green) and $k<0$ (blue). The simulation has been done either with noise (empty circles)
  or without noise (half-filled circles). The $\alpha$ values are given by the
  OLS estimator. The
  horizontal violet line is the statistical uncertainty on $\alpha$,
  the shadowed region under it is thus the non-significance zone of $\alpha$ values. (c): Allan deviation of the expected bias on the
  gravitational redshift of Eq. \ref{eq-redshift} at order 1:$-GM(1/r_k-1/r_{\mathrm{POD}})/c^2$ with
  $r_k$ the norm of the ISS position vector with the $k$-degraded orbitography. The legend indicates the error magnification. The smooth black line is the specification
of PHARAO. For $k=1$, we also show the Allan deviation obtained with a sinusoidal
orbit error with ISS orbital period 5400~s and amplitude
$\sqrt{2}\sigma_r \simeq 42$~cm (in blue), as well as the theoretical line going through its maxima,
scaling as $1/\tau$ (in cyan).}
\label{fig-orbito}
\end{figure}

Here we carry out the same numerical estimation for the clock redshift part, using the
gravitational redshift test software presented in this article. To assess numerically the orbitography
requirements, we simulate the ``imperfect knowledge of
orbitography'' as follows. The ISS orbitography file we used for the
tests presented until here is a precise version, labeled POD (Precise
Orbit Determination). We also had a less accurate version of this
orbitography, labeled OD (Orbit Determination), see \cite{Meynadier2018} for details. The difference OD-POD provides a realistic estimate of orbitography error. The difference of the norm of the position
vectors between these two files is shown on Figure \ref{fig-difforbito}. Its
standard deviation is $\sigma_r\sim30$~cm. For our test we simulated data using the POD file. Then
for the analysis, when constructing the GR model
(both substracted from the data and then used in the design matrix) we used
several versions of the orbitography:

\begin{itemize}
\item[$-$] the POD orbitography (this should lead to unbiased results); 
\item[$-$] the OD orbitography;
\item[$-$] $k$-degraded versions of the OD orbitography, where we magnify
 by a factor $k$ the difference between the OD and POD positions and
 velocities and add this orbit error to the POD version. This factor
 is applied on all coordinates. We tested $k=\pm10^n$
 with $n$ from 0 to 5.
\end{itemize}

The simulation and analysis were carried out for the OPMT ground station, for phase
data with LSMC. The statistical uncertainty on $\alpha$ is not affected by the orbitography version
we use for the analysis and is equal to
$\sigma_{\alpha}=2.9\times10^{-6}$. The $\alpha$ values given by the
OLS estimator are pictured on Fig. \ref{fig-alphaorbito}. As can be seen from these values, the $k$ factor required to lead to
a statistically significant bias on our data is $\pm10^3$.  In order to
know the bias itself on $\alpha$ from the orbit error, we repeated the same
analysis but on data simulated without noise. The results are
also pictured on Fig. \ref{fig-alphaorbito}. As expected, the bias increases linearly with the
$k$ value. It exceeds the ppm level only for $k\geqslant 10^3$, which is
consistent with the observation of a statistically significant value
in the case of noise. This
degradation corresponds to a standard deviation of the orbit error of 300~m.

Further insight can be gained by visualizing the Allan
deviation of the expected bias on the gravitational redshift. It is shown on
Figure \ref{fig-adevorberr}, together with the PHARAO clock
uncertainty. One can observe that they have different averaging
trends, and that only when $k\geqslant10^3$, the bias Allan deviation overcomes the sensitivity
limitation from the PHARAO clock until at least 12 days. This is
consistent with the significant fit results
obtained only for $k\geqslant 10^3$ on Fig. \ref{fig-alphaorbito}. The
bias Allan deviation has a bump at half the orbital
period, indicating that the error has a strong component at orbital
frequency. In order to check this behaviour and understand its averaging, we also plotted the Allan deviation of the
gravitational redshift bias expected from a sinusoidal orbitography
error with the orbital period and an amplitude $\sqrt{2}\sigma_r$. It
has minima due to its periodicity, between bumps that match the
position and height of the ones of the realistic deviation, which
confirms the periodic trend of the orbit error. The maxima of this model
decay as $1/\tau$, which is also plotted on the figure and reproduces well
the decay of the realistic signal.

Our conclusion is thus that the maximum allowed orbit error to fulfil the goal of the ACES redshift test at 2 ppm is
$\sim$300 m, which will be easily achieved since the present OD
version we have has only an error of 30~cm.

We note that more stringent requirements may be necessary for other science objectives. In particular if the degradation of the observed clock stability from orbit errors is required to stay below the PHARAO instability at all averaging times, then Figure \ref{fig-orbito} shows that $k$ needs to stay below $\approx$~80, corresponding to a maximum orbit error of 24~m, which is of the same order of magnitude as the estimate in \cite{Duchayne2009}.




\section{Conclusion and perspectives} \label{sec:Conclusion}

We have presented an end to end performance study of the ACES mission with respect to its primary scientific objective: testing of the gravitational redshift. We assume that the PHARAO systematic uncertainty of about 1 part in $10^{16}$ in fractional frequency will be the ultimate limit in the redshift test, providing a performance of 2 ppm, about an order of magnitude better than the state of the art. Under that assumption, we determine whether that limit can be reached in a realistic scenario concerning dead time, main noise sources, and orbit determination errors.

Our answer is ``yes'', when taking into account a realistic observation scenario (large, irregular data gaps making for $\sim$97\% dead time) and the ACES specifications for link and onboard clock noise. Remarkably, we find that the goal can be reached with only a few (min. 3) experimental sessions of 10-20 days, with a single ground station and with orbit determination errors as large as $\sim$300~m. Nonetheless, in practice several (up to 8) ground stations will participate in order to improve robustness and to allow identification of potential station dependent systematic biases. Also, our results only apply to the test of the gravitational redshift, other science objectives might require more measurement sessions and/or more ground stations and/or better orbit determination.

We have studied several statistical analysis methods and discussed their respective performance and merits. We have also described in some detail the software used for data simulation and data analysis. The latter is the one we intend to use once real ACES data is available.

Our study also has some implications for the design and definition of future missions like the STE-QUEST \cite{Altschul2015} or SOC \cite{Origlia2016} projects. In particular, we can use our software to quantitatively evaluate the performance in terms of the redshift test for any clock in an Earth-orbit scenario as a function of ground station distribution, noise levels and orbit determination errors. 




\newpage
\ack

We acknowledge valuable discussions with Anja Schlicht and Stefan Marz, which allowed us to
test the simulation software. We thank Oliver Montenbruck for
providing the ISS orbitography files.

\section*{References}
\bibliographystyle{iopart-num}
\bibliography{redshift}
\appendix
\section{Inverse of the random walk noise covariance matrix with
  data gaps}\label{sec-rw}
\setcounter{section}{1}
In order to perform the comparison between Generalized Least Squares (GLS) and Least Squares Monte Carlo (LSMC), we had to use the inverse of the covariance matrix for a random walk noise (RWN) in presence of data gaps. As easy as it is to numerically compute a covariance matrix, the huge number of data points ($n=10^6$) prevented us from computing the inverse of the RWN covariance matrix due to computational limitation (the storage and inversion of a $n\times n = 10^{12}$ elements matrix). Thanks
to the peculiarity of the RWN covariance matrix, we still manage to perform the test described in section \ref{sb-compGLS}.

\subsection{Random walk noise covariance matrix without data gaps} 
First, let us consider the case without any data gaps. 
With this hypothesis and considering $n$ regularly spaced data points (separated by $\Delta t$), the $(n,n)$ covariance matrix of a RWN $\Omega$ and its inverse $\Omega^{-1}$ can be simply represented with the following formal expression for an element $(i,j)$ of each matrix:
\begin{equation}
\boxed{
\begin{array}{c}
\Omega_{i,j} = \sigma^2 \Delta t^2 \textsc{min}\left(i,j\right) \\
\\
\left(\Omega^{-1}\right)_{i,j} = \frac{1}{\sigma^2\Delta t^2}\left\{
\begin{array}{cc}
1 &\text{if} \quad i = j = n \\
-1 &\text{if} \quad  i = j \pm 1 \\
2 & \text{if} \quad i = j \neq n\\
0 & \text{elsewhere}
\end{array}
\right.
\end{array}}
\end{equation}
where $\sigma$ is the standard deviation of the underlying white noise
that was integrated to obtain the RWN. In the following, we will consider $\sigma =1$.
\begin{figure}[h!]
\begin{center}
\includegraphics[scale=0.4]{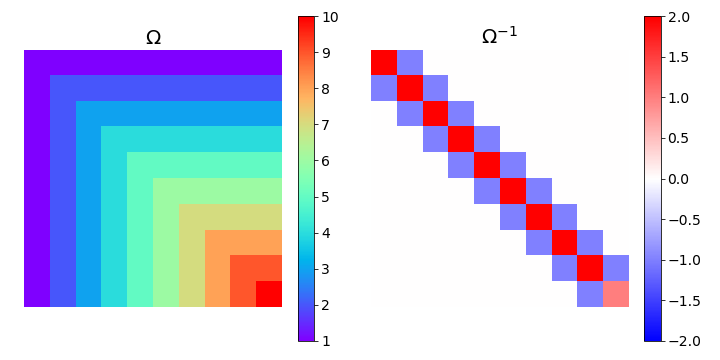}
\end{center}
\vspace{-.5 cm}
\caption{Example of $\Omega$ and $\Omega^{-1}$ for a rank $10$ matrix and a random walk noise}
\end{figure}

As we can see, the inverse matrix $\Omega^{-1}$ is tri-diagonal. We shrunk down the number of relevant elements in $\Omega^{-1}$ from $n\times n \simeq 10^{12}$ to $3\times n = 3 \times 10^6$. This result is important since it enables us to predict analytically and store the inverse covariance matrix.
The next step is to consider the effect of data gaps on this result.

\subsection{Random walk noise covariance matrix with data gaps} 
The data gaps in the software can be represented as a $(m,n)$ mask array $M$ where $m$ is the number of remaining data points. The effect of $M$ is to remove the rows and columns of the original matrix that correspond to data gaps. The matrix $M \Omega M^T$ will be the RWN covariance matrix of the data with gaps :
\begin{equation}
M \Omega M^T = \left( 
\begin{array}{c c c c c}
q_1 & q_1 & \hdots & \hdots & q_1 \\
q_1 & q_2 & \hdots & \hdots & q_2 \\
\vdots & \vdots & \ddots & & \vdots \\
\vdots & \vdots & & \ddots & q_{m-1}\\
q_1 & q_2 & \hdots & q_{m-1} & q_{m}
\end{array}\right) \qquad \text{with} \quad q_i \in \mathbb{Z}.
\end{equation}
As we can see, the matrix structure is preserved and therefore, we can find an analytical formula for its inverse.

\subsection{Inverse of the random walk noise covariance matrix with data gaps}
Using the tridiagonal matrix inversion recurrence relations, we have the value of each element $(i,j)$ of the RWN covariance matrix inverse :
\begin{equation}
\boxed{
\left(\left(M\Omega M^T\right)^{-1}\right)_{i,j} = \left\{
\begin{array}{cc}
\Delta_{m-1} & \text{if} \quad i = j = m \\
\frac{1}{q_1} + \Delta_1 & \text{if} \quad i = j = 1 \\
\Delta_{i-1} + \Delta_{i} & \text{if} \quad i = j \neq m,1 \\
- \Delta_{i} & \text{if} \quad i = j - 1 \\
- \Delta_{j} & \text{if} \quad i = j + 1 \\
0 & \text{elsewhere}
\end{array}
\quad \text{with} \quad \Delta_i = \frac{1}{q_{i+1}-q_{i}}.\right.}
\end{equation}

\begin{figure}[h!]
\begin{center}
\includegraphics[scale=0.4]{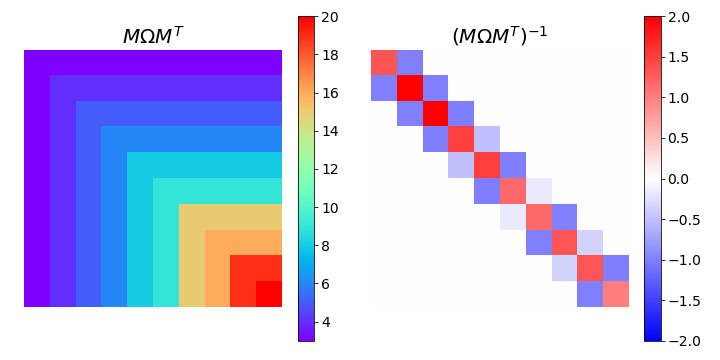}
\end{center}
\vspace{-.5 cm}
\caption{$M \Omega M^T$ and $\left(M \Omega M^T\right)^{-1}$ with random gaps}
\end{figure}

\noindent This formula represents a tridiagonal matrix that can easely be stored in a laptop computer and is the final product needed to perform the GLS.

\section{Inverse of the violet noise covariance matrix with data gaps}\label{sec-vn}
Another remarkable result applies to the derivative of a white noise, commonly referred to as violet noise (VN). 

\subsection{Violet noise covariance matrix and its inverse without data gaps} 
At first, let us consider the case without any data gaps. 
The demonstration of the covariance matrix for a violet noise relies on the white noise definition. Let $Y$ be a random variable affected by white noise and $X$ its derivative, $X = \frac{d Y}{d t}$, so that $X$ is affected by VN.
An element $(i,j)$ of the VN covariance matrix $\Omega$ is :
\begin{equation}
\Omega_{i,j} = \mathrm{Cov}\left[x_i,x_j\right] = \mathrm{Cov}\left[\frac{d y_i}{d t},\frac{d y_j}{d t}\right] \simeq \mathrm{Cov}\left[\frac{y_i - y_{i-1}}{\Delta t},\frac{y_j - y_{j-1}}{\Delta t}\right] = \frac{1}{\Delta t^2} \mathrm{Cov}\left[y_i - y_{i-1},y_j - y_{j-1}\right]
\end{equation}
The above expression can be reduced when using the bilinearity of the covariance and the covariance of a white noise $\mathrm{Cov}\left[y_i,y_j\right] = \sigma^2 \delta_{(i-j)}$ where $\delta$ is the Kronecker delta. We also take advantage of the fact that $\Omega$ is tridiagonal to get $\Omega^{-1}$ :
\begin{equation}
\boxed{
\begin{array}{c}
\Omega_{i,j} = \frac{\sigma^2}{\Delta t^2} \left[ 2\delta_{(i-j)} -\delta_{(|i-j|-1)}\right]
\\
\\
\left(\Omega^{-1}\right)_{i,j} = \frac{\Delta t^2}{\sigma^2}
\left\{ \begin{array}{c c c}
\frac{i \left(n+1-j\right)}{n+1} & & i \leq j \\
\frac{j \left(n+1-i\right)}{n+1} & & i > j \\
\end{array} \right.
\end{array}
}
\end{equation}
where $\sigma$ is the noise standard deviation. In the following, we will consider $\sigma/\Delta t =1$.
\begin{figure}[h!]
\begin{center}
\includegraphics[scale=0.4]{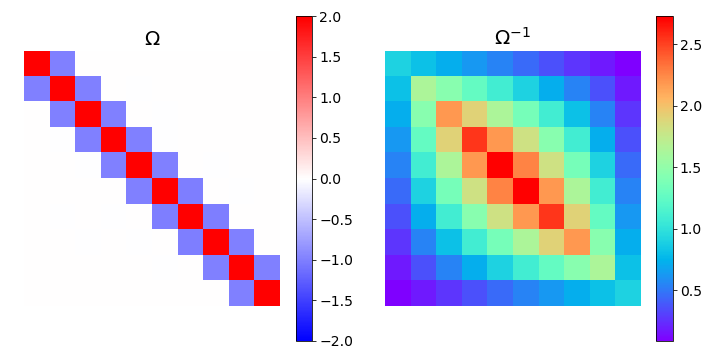}
\end{center}
\vspace{-.5 cm}
\caption{Example of $\Omega$ and $\Omega^{-1}$ for a rank $10$ matrix and violet noise}
\end{figure}

\subsection{Violet noise covariance matrix with data gaps} 
Using the same mask array as in the previous section, the VN covariance matrix becomes a block diagonal matrix where the sub-matrices are smaller versions of the full VN covariance matrix.
Let $\Omega$ be a $(n,n)$ VN covariance matrix and $M$ the mask array. $M$ will create submatrixes of size ${m_1,m_2,...,m_p}$ where $m_i$ is the length of the uninterrupted data before the $i$-th gap and $p$ is the number of gaps.
The effect of the mask $M$ on the full VN covariance matrix $\Omega$ is :
\begin{equation}
M \Omega M^T= 
 \left( 
\begin{array}{c c c c}
\Omega_{m_1} & & &0 \\
 & \Omega_{m_2}& & \\
 & & \ddots & \\
0 & & & \Omega_{m_p} \\
\end{array} 
\right)
\end{equation}

\subsection{Inverse of the VN covariance matrix with data gaps} 
The masked VN covariance matrix is block diagonal and the inversion is simply the inverse of each submatrix placed on the full matrix diagonal:
\begin{equation}
\left(M \Omega M^T\right)^{-1}	= \left( 
\begin{array}{c c c c}
\Omega_{m_p}^{-1} & & & 0\\
 & \Omega_{m_2}^{-1} & & \\
 & & \ddots & \\
0 & & & \Omega_{m_p}^{-1} \\
\end{array} 
\right)
\end{equation}
\begin{figure}[h!]
\begin{center}
\includegraphics[scale=0.4]{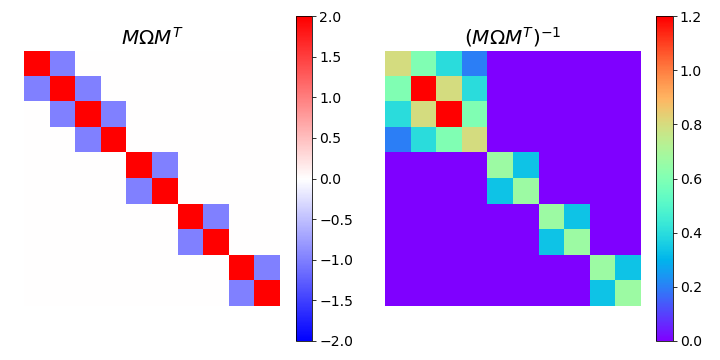}
\end{center}
\vspace{-.5 cm}
\caption{$M \Omega M^T$ and $\left(M \Omega M^T\right)^{-1}$ with random gaps}
\end{figure}

\noindent The VN covariance matrix inverse is not necessarily sparse since it is highly dependant on the data gaps. Basically, the size of the matrix goes from $n \times n$ to $\Sigma_{i} m_i^2$.
For the ACES-PHARAO mission, the inverse of the covariance matrix becomes sparse when we consider the large gaps in the data. This means, again, that the final product can be stored and used to perform the GLS.

\end{document}